\documentclass{aastex}
\usepackage{emulateapj5,times,mathptm}

\newcommand {\asec} {$^{\prime\prime}$}

\newcommand{\Lsolar}{\mbox{\,$\rm L_{\odot}$}}

\def\amin{\ifmmode ^{\prime}\else$^{\prime}$\fi}
\def\asec{\ifmmode ^{\prime\prime}\else$^{\prime\prime}$\fi}

\def\etal{{et\,al.\,}}

\def\asca{{\it ASCA\/}}

\def\chandra{{\it Chandra\/}}

\def\heao1{{\it HEAO-1\/}}

\def\rosat{{\it ROSAT\/}}
\def\scuba{{SCUBA\/}}

\def\xmm{{\it XMM-Newton\/}}

\def\ltsima{$\; \buildrel < \over \sim \;$}
\def\simlt{\lower.5ex\hbox{\ltsima}}
\def\gtsima{$\; \buildrel > \over \sim \;$}
\def\simgt{\lower.5ex\hbox{\gtsima}}

\journalinfo{The Astronomical Journal, 2003 February, astro-ph/0211267}

\begin{document}
%

%
\title{The Chandra Deep Field-North Survey. XIV. X-ray detected Obscured AGNs and Starburst Galaxies in the Bright Submm Source Population}
%

\author{D.M.~Alexander, F.E.~Bauer, W.N.~Brandt, A.E.~Hornschemeier, C.~Vignali, G.P.~Garmire, D.P.~Schneider, G.~Chartas and S.C.~Gallagher}

\affil{Department of Astronomy \& Astrophysics, 525 Davey Laboratory, The Pennsylvania State University, University Park, PA 16802}


\shorttitle{OBSCURED AGNS AND STARBURST GALAXIES IN THE BRIGHT SUBMM SOURCE POPULATION}

\shortauthors{ALEXANDER ET AL.}

\slugcomment{Received 2002 July 3; accepted 2002 Nov 12}

%
\begin{abstract}
%

We provide X-ray constraints and perform the first X-ray spectral analyses for bright \scuba\ sources ($f_{\rm 850\mu m}\ge$~5~mJy; S/N$\ge4$) in an $8\farcm4\times8\farcm4$ area of the 2~Ms \chandra\ Deep Field-North survey containing the Hubble Deep Field-North. X-ray emission is detected from seven of the ten bright submm sources in this region down to 0.5--8.0~keV fluxes of $\approx1\times10^{-16}$~erg~cm$^{-2}$~s$^{-1}$, corresponding to an X-ray detected submm source density of $360^{+190}_{-130}$~deg$^{-2}$; our analyses suggest that this equates to an X-ray detected fraction of the bright submm source population of $\simgt$~36\%, although systematic effects may be present. Two of the X-ray detected sources have nearby (within 3\asec) X-ray companions, suggesting merging/interacting sources or gravitational lensing effects, and three of the X-ray detected sources lie within the approximate extent of the proto-cluster candidate \hbox{CXOHDFN J123620.0+621554}. Five of the X-ray detected sources have flat effective X-ray spectral slopes ($\Gamma<1.0$), suggesting obscured AGN activity. X-ray spectral analyses suggest that one of these AGNs may be a Compton-thick source; of the other four AGNs, three appear to be Compton-thin sources and one has poor constraints. The rest-frame unabsorbed X-ray luminosities of these AGNs are more consistent with those of Seyfert galaxies than QSOs (i.e.,\ $L_{\rm X}\approx 10^{43}$--$10^{44}$~erg~s$^{-1}$). Thus, the low X-ray detection rate of bright submm sources by moderately deep X-ray surveys appears to be due to the relatively low luminosities of the AGNs in these sources rather than Compton-thick absorption. A comparison of these sources to the well-studied heavily obscured AGN NGC~6240 shows that the average AGN contribution is negligible at submm wavelengths. The X-ray properties of the other two X-ray detected sources are consistent with those expected from luminous star formation; however, we cannot rule out the possibility that low-luminosity AGNs are present. The three X-ray undetected sources appear to lie at high redshift ($z>4$) and could be either AGNs or starbust galaxies.

\end{abstract}


%
\section{Introduction}\label{intro}
%

Due to the strong negative $K$-correction for galaxies at submillimeter (submm; $\lambda$~=300--1000~$\mu$m) wavelengths, submm surveys facilitate observations of the high-redshift universe (e.g.,\ Blain \& Longair 1993). For typical starburst galaxies and Active Galactic Nuclei (AGNs), the expected submm flux density is approximately insensitive to redshift for $z\approx$~1--10. Indeed, the majority of the submm-detected sources in \scuba\ (Holland \etal 1999) surveys appear to be extremely luminous infrared galaxies at $z\approx$~1--6 (e.g., Smail, Ivison, \& Blain 1997; Hughes \etal 1998; Ivison \etal 1998; Smail \etal 2000, 2002; Dunlop 2001; Blain \etal 2002).

It is generally believed that these submm-detected sources are proto-galaxies undergoing intense dust-enshrouded star formation (e.g.,\ Blain \etal 1999; Lilly \etal 1999). However, evidence for AGNs has also been found using a variety of techniques. For example, optical spectroscopic observations have revealed AGNs in a number of optically bright submm sources (e.g.,\ Ivison \etal 1998; Barger \etal 1999a; Soucail \etal 1999), and \scuba\ observations of known optically bright Quasi-Stellar Objects (QSOs; i.e.,\ luminous type 1 AGN) have shown that luminous AGNs comprise a fraction of the submm source population (e.g.,\ McMahon \etal 1999; Page \etal 2001; Isaak \etal 2002; Willott \etal 2002).\footnote{The exact fraction of optically bright QSOs in the submm source population is poorly known. Since only one of the 15 sources in the \scuba-lens survey (Smail \etal 2002) is a QSO [SMM J02399--0136 is a broad absorption line QSO (BALQSO); see Vernet \& Cimatti 2001], it is probably of the order of $\approx$~5--10\%.} Although clearly some \scuba\ sources host AGNs, the ubiquity of AGNs in the submm source population, and their contribution to the submm emission, are poorly constrained.

Arguably the best discriminator of AGN activity is the detection of hard X-ray emission (i.e.,\ $>$~2~keV). Hard X-ray observations are particularly useful in identifying AGNs in sources where the optical signatures are weak (e.g.,\ optically faint AGNs or AGNs without strong emission lines; Vignati \etal 1999; Mushotzky \etal 2000; Alexander \etal 2001; Hornschemeier \etal 2001). Due to their high-energy X-ray coverage, high X-ray sensitivity, and excellent positional accuracy, the \chandra\ (Weisskopf \etal 2000) and \xmm\ (Jansen \etal 2001) observatories offer the best opportunities for the X-ray investigation of \scuba\ sources. Although sensitive enough to detect luminous AGNs out to high redshift, the cross-correlation of moderately deep X-ray surveys with \scuba\ surveys has yielded little overlap between the X-ray and submm detected source populations (Fabian \etal 2000; Hornschemeier \etal 2000, 2001; Severgnini \etal 2000; Barger \etal 2001a; Almaini \etal 2002; however, see also Bautz \etal 2000 and Ivison \etal 2002). These studies have suggested that a large number of bolometrically important AGNs (i.e.,\ AGNs that can dominate the submm emission) can only be present in the submm source population if they are obscured by Compton-thick (i.e.,\ $N_{\rm H}> 1.5\times 10^{24}$~cm$^{-2}$) material.

%
%

\begin{figure*}[t]
\centerline{\includegraphics[width=19.5cm]{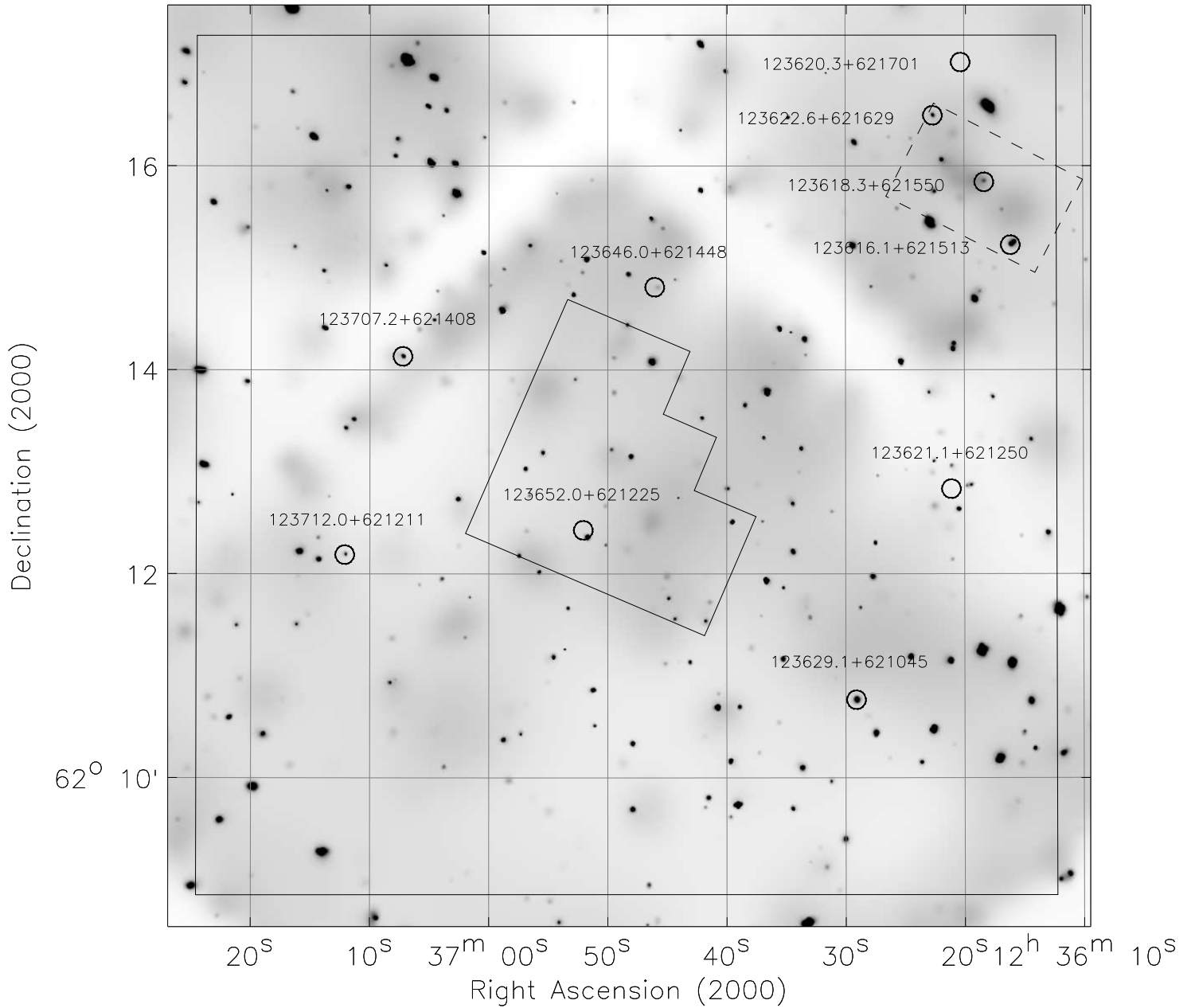}}
\vspace{0.1in}
\figcaption{Adaptively smoothed full-band 2~Ms \chandra\ image. This image has been made using the standard \asca\ grade set and has been adaptively smoothed at the $2.5\sigma$ level using the code of Ebeling, White, \& Rangarajan (2002). The HDF-N is shown as the polygon at the center of the image, and the large box indicates the reduced Hawaii flanking-field area used in this study. The circles indicate the source positions of the labelled submm sources (see \S2.2 and Table~1). The dashed box indicates the position and approximate extent of CXOHDFN J123620.0+621554 (e.g.,\ Bauer \etal 2002a), a candidate high-redshift cluster or proto-cluster that has an overdensity of submm sources (see \S2.4.1); since this source emits predominantly in the soft band it is not apparent in this image. Most of the faint extended emission is instrumental background or artifacts of the smoothing algorithm. The two white streaks towards the top of the image are due to gaps in the ACIS-I CCDs. The faintest X-ray sources are below the significance level of the smoothing and thus are not visible in this figure.}
\label{sourceexp}
\vspace{0.2in}
\end{figure*}


Targeting X-ray sources detected in the 1~Ms \chandra\ Deep Field-North (CDF-N; Brandt \etal 2001) survey with \scuba\ jiggle-map observations, Barger \etal (2001b) showed that submm counterparts can be found for faint X-ray sources. Five of the six X-ray detected submm sources with good X-ray constraints had flat ($\Gamma<1.0$) effective X-ray spectral slopes, an indicator of obscured AGN activity. However, while clearly showing that many of the AGNs found in submm sources are obscured, Barger \etal (2001b) did not investigate the possibility that the absorption toward the AGNs is Compton thick. Thus strong constraints on the origin of the submm emission were not placed.

The primary aim of this paper is to place strong constraints on the origin of the submm emission from AGN-classified submm sources by determining if their AGNs are obscured by Compton-thin or Compton-thick material. In this study we focus on an $8\farcm4\times8\farcm4$ region within the 2~Ms CDF-N survey with excellent multi-wavelength data, including deep radio (Richards \etal 1998; Richards 2000) and optical/near-infrared (near-IR) observations (Barger \etal 1999b). Due to the extremely high sensitivity of the 2~Ms \hbox{CDF-N} survey, we have the potential to detect X-ray emission from high-luminosity starburst galaxies (i.e., $L_{\rm X}\approx 10^{43}$~erg~s$^{-1}$) to $z\approx$~2.5. In this study we place the tightest constraints to date on the X-ray emission from star formation in submm-detected galaxies. The Galactic column density along this line of sight is $(1.6\pm 0.4)\times 10^{20}$~cm$^{-2}$ (Stark \etal 1992), and $H_{0}=65$~km~s$^{-1}$~Mpc$^{-1}$, $\Omega_{\rm M}=\onethird$, and $\Omega_{\Lambda}=\twothirds$ are adopted throughout.

%
\section{Observations and Source Samples}\label{data}
%

\subsection{X-ray Observations}

The X-ray results reported in this paper were obtained with ACIS-I (the imaging array of the Advanced CCD Imaging Spectrometer; Garmire et~al. 2002) onboard \chandra. The 2~Ms CDF-N observations were centered on the Hubble Deep Field-North (HDF-N; Williams \etal 1996) and cover $\approx460$~arcmin$^2$. The region investigated in this study extends over an $8\farcm4\times8\farcm4$ area that has its center near to the aim-point of the CDF-N observations. This 70.3~arcmin$^2$ region is defined by the optical and near-IR observations of Barger \etal (1999b) and is referred to here as the reduced Hawaii flanking-field area (see Alexander \etal 2001). The X-ray data processing of these observations was similar to that described in Brandt \etal (2001) for the 1~Ms \chandra\ exposure. The full 2~Ms source catalog is presented in Alexander \etal (2003).

One hundred and ninety-three (193) X-ray sources are detected in the reduced Hawaii flanking-field area with a {\sc wavdetect} (Freeman \etal 2002) false-positive probability threshold of 10$^{-7}$ down to on-axis 0.5--2.0~keV (soft-band) and 2--8~keV (hard-band) flux limits of $\approx1.5\times10^{-17}$~erg~cm$^{-2}$~s$^{-1}$ and $\approx1.0\times10^{-16}$~erg~cm$^{-2}$~s$^{-1}$, respectively. The X-ray coverage in this region is fairly uniform, with 0.5--8.0~keV (full-band) effective exposure times of the 193 X-ray detected sources covering 1.28--1.94~Ms (with a median time of 1.82~Ms). The adaptively smoothed full-band 2~Ms \chandra\ image is shown in Figure~1. The positional uncertainties of the X-ray sources are $\simlt$~1\farcs0, depending on off-axis angle (see Alexander \etal 2003), and the median positional uncertainty is 0\farcs4. These small positional uncertainties allow for accurate cross-correlation with multi-wavelength counterparts.

\subsection{Bright Submm Sources}

In this study we focus on bright submm ($f_{\rm 850\mu m}\ge$~5~mJy) sources detected with a signal-to-noise ratio (S/N) $\ge$~4. This threshold was chosen to guard against possible spurious \scuba\ sources (e.g.,\ see the estimated number of spurious sources at different significance thresholds for the \scuba\ observations of the ELAIS-N2 and Lockman Hole East regions; Scott \etal 2002).

The submm data used here was taken from a number of different studies (Hughes \etal 1998; Barger \etal 2000, 2002; Borys \etal 2002). Ten bright submm sources are detected with S/N$\ge$~4; see Table~1 and Figure~1.\footnote{We also included the studies of Chapman \etal (2001) and Serjeant \etal (2002) but no further \scuba\ sources met our criteria.} If we relaxed the detection threshold to S/N$\ge$~3.5 another five sources would be added to our sample; however, $\approx$~20\% of the sources in the total sample could be spurious (e.g.,\ Scott \etal 2002). Although we have utilized all of the published \scuba\ data currently available, this region only has complete \scuba\ coverage down to $f_{\rm 850\mu m}=$~12~mJy ($\approx$~75\% of the region has \scuba\ coverage down to $f_{\rm 850\mu m}=$~5~mJy; e.g.,\ see Figure 1 of Barger \etal 2002). Therefore, there are likely to be further bright submm sources in this region that have not yet been detected by \scuba. 

Two main observing techniques were employed in these studies: Hughes \etal (1998) and Borys \etal (2002) performed \scuba\ scan-map observations while Barger \etal (2000, 2002) performed \scuba\ jiggle-map observations of radio and X-ray sources. The positional uncertainty of a \scuba\ source is dependent on the signal-to-noise ratio and is typically $\approx$~2--4\asec\ (e.g., Serjeant \etal 2002; Smail \etal 2002; although see Hogg 2001). However, in the cases where radio and X-ray sources have been targeted with \scuba\ jiggle-map observations, $<$1\asec\ positions can be inferred from the radio and X-ray data. 

Seven of the ten submm sources were detected in \scuba\ jiggle-map observations of radio and X-ray sources, and have positional uncertainties of $<$1\asec. For these sources we identified optical and near-IR counterparts in the photometric source catalogs produced by Alexander \etal (2001) using a 1\asec\ matching radius (see Table~1); these source catalogs were produced from the Barger \etal (1999b) $I$-band and $HK^{\prime}$-band images which reach $\approx2\sigma$ magnitude limits of $I=25.3$ and $HK^{\prime}=21.4$, respectively.\footnote{These images are publicly available at http://www.ifa.hawaii.edu/$\sim$cowie/hdflank/hdflank.html. The relationship between the $K$ band and $HK^\prime$ band is $HK^\prime-K=0.13+0.05(I-K)$ (Barger \etal 1999b).}

Three of the ten submm sources were detected in \scuba\ scan maps and have positional uncertainties of $\approx$~2--4\asec. However, in the case of one source (123652.0+621225, which is often referred to as HDF850.1, the brightest submm source in the HDF-N; Hughes \etal 1998) we can adopt the sub-arcsecond 1.3~mm source position obtained with the IRAM Interferometer on the Plateau de Bure (IRAM PdBI) telescope (i.e.,\ Downes \etal 1999). Dunlop \etal (2002) used these data in combination with extremely sensitive MERLIN+VLA 1.4~GHz radio observations to identify accurately the optical and near-IR counterpart for 123652.0+621225. Since this source is located in the HDF-N, the optical and near-IR constraints are more sensitive than for the other submm sources investigated here (123652.0+621225 has $I>28.6$ and $K=23.4$; Dunlop \etal 2002). We must rely on the \scuba\ positions for the other two sources since neither source has a radio or X-ray counterpart within 4\asec. The $\approx$~2--4\asec\ positional uncertainties for these sources are too large to be able to identify optical and near-IR counterparts relibably.

\subsection{Source Redshifts}

Only one of the submm sources has an optical spectroscopic redshift (123629.1+621045 has $z=1.013$; Cohen \etal 2000; Hornschemeier \etal 2001) and another source has a redshift estimate based on its multi-wavelength spectral energy distribution (SED; 123652.0+621225 has $z=4.1^{+0.5}_{-0.6}$; Dunlop \etal 2002). However, on the assumption that the other sources obey the radio-to-far-infrared (far-IR; $\lambda$~=40--120~$\mu$m) correlation found for local star-forming galaxies (e.g.,\ Helou, Soifer, \& Rowan-Robinson 1985), we can estimate redshifts for all of the radio-detected sources, and place redshift lower limits on the radio-undetected sources, using the millimetric redshift technique (e.g., Carilli \& Yun 1999). Since the submm emission corresponds to the longer wavelength Rayleigh-Jeans tail of the thermal dust emission detected at far-IR wavelengths, redshifts are directly determined (albeit with considerable uncertainty) on the basis of the radio-to-submm spectral slope. As noted by Smail \etal (2000), this technique should give a lower limit on the source redshift when a component of the radio emission is produced by an AGN.

We determined millimetric redshifts using the redshift estimators of Carilli \& Yun (2000), Eales \etal (2000), and Rengarajan \& Takeuchi (2001). We found that the Eales \etal (2000) estimator gave the lowest redshift determinations (a minimum redshift of $z=1.3$ and a maximum redshift of $z>3.2$), while the Rengarajan \& Takeuchi (2001) estimator gave the highest redshift determinations (a minimum redshift of $z=2.0$ and a maximum redshift of $z>4.8$). The differences in the redshift determinations are due to the assumed SED and the temperature of the dust that produces the far-IR-to-submm emission. The millimetric redshifts of the source with an optical spectroscopic redshift (123629.1+621045; $z=1.013$) cover $z=$~1.4--2.3 ($z=$~1.0--3.0 when the 1$\sigma$ millimetric redshift uncertainties are taken into account). Clearly the optical spectroscopic redshift of 123629.1+621045 is marginally in agreement with the millimetric redshift estimates (we note that larger studies find a $\simgt$~50\% agreement between the optical spectroscopic redshifts and millimetric redshifts of submm sources; e.g.,\ Smail \etal 2000, 2002). However, in the absence of better redshift estimations we must determine the redshifts for the majority of these sources using the millimetric redshift technique.

The adopted millimetric redshifts for our sources are determined using the Carilli \& Yun (2000) redshift estimator, which provides a good compromise between the Eales \etal (2000) and Rengarajan \& Takeuchi (2001) millimetric redshift estimates (see Table 1). We note that since the typical 1$\sigma$ millimetric redshift uncertainty for a source is $\Delta$$z=$~0.8 (the range of redshift uncertainties is $\Delta$$z=$~0.3--1.1), the differences in these redshift determinations are not critical. However, even given these uncertainties, all sources have $z>1$ with all millimetric redshift estimators.

\subsection{X-ray Detected Submm Sources}

Five of the ten submm sources were detected by Barger \etal (2002) in \scuba\ jiggle-map observations of X-ray sources and are thus X-ray detected submm sources (see Table~2). We searched for X-ray counterparts for the other five submm sources using a 1\asec\ matching radius for the three sources with a radio counterpart and a 4\asec\ matching radius for the two sources without a radio counterpart.\footnote{X-ray sources detected in the soft band (0.5--2.0~keV), hard band (2--8~~keV) or full band (0.5--8.0~keV) were matched to submm counterparts.} One further X-ray counterpart was found for one of the radio-detected submm sources (123622.6+621629; see Table~2). 

Given the low surface density of the submm sources, we can also search for lower significance X-ray sources associated with submm sources without introducing a significant number of spurious X-ray sources. We ran {\sc wavdetect} with a false-positive probability threshold of 10$^{-5}$ and found one further match to a radio-detected submm source (123618.3+621551; see Table~2); the probability of this X-ray-submm match being spurious is \hbox{$<1$\%}. This source was also detected in the 1~Ms \chandra\ exposure at a 10$^{-5}$ false-positive probability threshold (Alexander \etal 2001) and is probably not detected here at a higher probability threshold since it lies within the bright extended X-ray source \hbox{CXOHDFN J123620.0+621554} (e.g.,\ Bauer \etal 2002a). Thus, seven ($70^{+30}_{-26}$\%) of the ten submm sources have X-ray counterparts.\footnote{All errors are taken from Tables 1 and 2 of Gehrels (1986) and correspond to the $1\sigma$ level; these were calculated assuming Poisson statistics.} 

Three submm sources are X-ray undetected. There is no evidence for excess X-ray emission in the \chandra\ images for two of these sources (123620.3+621701 and 123621.1+621250); however, there is a suggestion of soft-band emission at the location of 123652.0+621225. Indeed, 123652.0+621225 lies within 0\farcs2 of the position of an X-ray source detected by {\sc wavdetect} at a 10$^{-4}$ false-positive probability threshold in the 1~Ms CDF-N study of Very Red Objects (VROs, $I-K\ge4$; Alexander \etal 2002b). Since 123652.0+621225 lies closer to the X-ray source than the VRO reported in Alexander \etal (2002b), it may actually be the correct counterpart to this extremely faint low-significance X-ray source. However, since the X-ray source is not detected at a higher significance level in the 2~Ms \chandra\ exposure, we consider 123652.0+621225 undetected here.

Examining the association between \chandra\ and \scuba\ sources, Almaini \etal (2002) found little overlap between the source populations but detected a strong clustering signal. They suggested that \chandra\ and \scuba\ sources are generally unrelated (at least down to the depths of their 75~ks \chandra\ survey) but trace the same large-scale structure. Although there may be some large-scale structure effects present in our sample, the cross-correlation distances between our \chandra\ and \scuba\ sources are much smaller than the up to 100~\asec\ clustering distances found by Almaini \etal (2002). Thus, we expect all of our X-ray-submm matches to be legitimate.

%
%
\begin{figure*}[t]
\centerline{\includegraphics[angle=0,width=12.0cm]{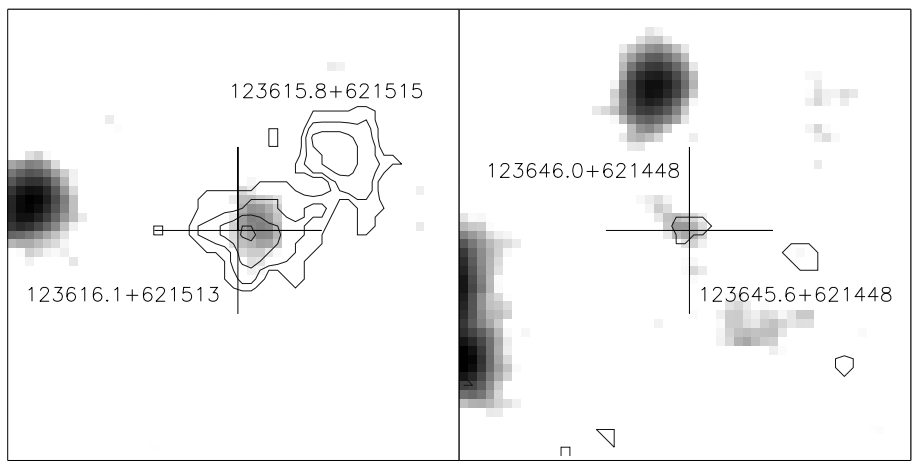}}
\figcaption{$I$-band images of the X-ray detected submm sources with nearby X-ray companions. The contours indicate the extent and intensity of the X-ray emission; the lowest contour corresponds to 2 counts pixel$^{-1}$ and each successive contour represents a factor of two increase in the number of counts pixel$^{-1}$. The large crosses mark the locations of the 1.4~GHz radio counterparts and the assumed submm counterparts. Each image is 9\farcs43 on a side.}
\label{fig:redshift1}
\vspace{0.2in}
\end{figure*}


\subsubsection{X-ray Detected Submm Source Density}

The derived density of X-ray detected submm sources in this region with $f_{\rm 850\mu m}\ge$~5~mJy is $360^{+190}_{-130}$~deg$^{-2}$. This source density should be considered a lower limit since there may be further bright submm sources with X-ray counterparts in this region that have not yet been detected by \scuba\ (see \S2.2).

We can compare our source density to that found in other studies. The \scuba\ survey of the ELAIS-N2 and Lockman Hole East regions (i.e.,\ Scott \etal 2002) contains sensitive \chandra\ and \xmm\ observations (down to 0.5--8.0~keV flux limits of $\approx10^{-15}$~erg~cm$^{-2}$~s$^{-1}$; i.e.,\ Manners \etal 2002; Hasinger \etal 2001). One \chandra\ counterpart was matched to a submm source in the ELAIS-N2 region (Almaini \etal 2002), corresponding to an X-ray detected submm source density of $35^{+80}_{-30}$~deg$^{-2}$. Four \xmm\ counterparts were found for submm sources in the Lockman Hole East region (Ivison \etal 2002), corresponding to a higher X-ray detected submm source density of $120^{+90}_{-50}$~deg$^{-2}$. The X-ray detected submm source densities found in our study are $\approx$~3--10 times higher than those found by Almaini \etal (2002) and Ivison \etal (2002) due to the greater sensitivity of the 2~Ms \chandra\ observations.

Three X-ray detected submm sources lie within 1\amin\ of the X-ray source \hbox{CXOHDFN J123620.0+621554} (see Figure~1). This extended X-ray source could be a high-redshift \hbox{($z\approx$~1--2)} cluster or proto-cluster (Bauer \etal 2002a; Chapman \etal 2002b), and the similar millimetric redshifts of the three nearby X-ray detected submm sources suggest that they may be directly associated with CXOHDFN J123620.0+621554. The statistics are extremely poor on how common such sources are. However, if they are rare, then our X-ray detected submm source density is likely to be higher than average at this X-ray depth; thus our sample may be affected by ``cosmic variance''.

\subsubsection{Fraction of Submm Sources with X-ray Counterparts}

The reduced Hawaii flanking-field area does not have complete \scuba\ coverage down to $f_{\rm 850\mu m}=$~5~mJy (see \S2.2), and hence we cannot directly place constraints on the fraction of the bright submm source population with X-ray counterparts. However, based on the submm source density results of Eales \etal (2000), and the compilations of Blain \etal (2002), Cowie, Barger, \& Kneib (2002), and Smail \etal (2002), we would expect 300--1000 submm sources per deg$^2$ with $f_{\rm 850\mu m}\ge$~5~mJy. This suggests that the fraction of bright submm sources in this region with X-ray counterparts is likely to be $\simgt36$\%. Note that this source fraction was not directly measured in our field and systematic effects could be present.

\subsubsection{Nearby X-ray Companions}

Two of the X-ray detected submm sources (123616.1+621513 and 123646.0+621448) have X-ray counterparts within 3\asec\ (see Figure~2; Alexander \etal 2003). In both cases one X-ray source has a radio counterpart and one does not. In addition, the radio-detected sources have $I\approx$~24--25 while the radio-undetected sources are optically blank. Given the density of X-ray sources in this region, the probability of two X-ray sources lying within 3\asec\ of each other is $\simlt$~1\% and therefore these associations appear to be significant.

There are two obvious scenarios that can explain the close proximity of the sources in these two pairs: (1) these sources are merging or interacting, or (2) these sources are gravitationally lensed. Unfortunately, since these sources are optically faint, we cannot search for the features of merging, interacting, or gravitational lensing in the current optical images. However, in the gravitational-lensing scenario, each member of each pair would generally have similar fluxes at all wavelengths (e.g.,\ see review by Blandford \& Narayan 1992). Therefore, the fact that one member of each pair is detected at both optical and radio wavelengths, while the other is not, provides some evidence against gravitational lensing.\footnote{The deep and high-spatial resolution {\it Hubble Space Telescope (HST)} Advanced Camera for Surveys (ACS) observations to be taken in this field as part of the Great Observatories Origins Deep Survey (GOODS) project should provide a direct test of these two scenarios. See http://www.stsci.edu/science/goods/ for details of the GOODS project.}

In both cases we have identified the radio-detected X-ray source as the counterpart to the submm source since the radio and submm emission are more closely matched in frequency space than the X-ray and submm emission. However, it is possible that the radio-undetected X-ray sources could contribute to a fraction of the submm emission.

%
\section{X-ray Properties of Submm-detected Sources}\label{sx}
%

\subsection{Basic Source Properties}

We show the submm flux density versus full-band X-ray flux of all the sources in Figure~3. As a comparison we also show the submm-detected \rosat-selected optically bright QSOs (i.e.,\ luminous type 1 AGNs) of Page \etal (2001). There is a clear distinction in X-ray flux between the QSOs and our X-ray detected submm sources. The former would be detectable in $\simlt10$~ks \chandra\ exposures, while the majority of the latter are only detectable in $\simgt$100~ks \chandra\ exposures. Since optical spectroscopic observations of submm sources have shown that optically bright QSOs probably comprise only a small fraction of the submm source population (i.e.,\ $\approx$~5--10\%, see footnote 1; e.g.,\ Barger \etal 1999a; Ivison \etal 2000; Smail \etal 2002), $\simgt$100~ks \chandra\ exposures are clearly required to detect X-ray counterparts for typical bright submm sources. 

%
%
\centerline{\includegraphics[angle=0,width=9.0cm]{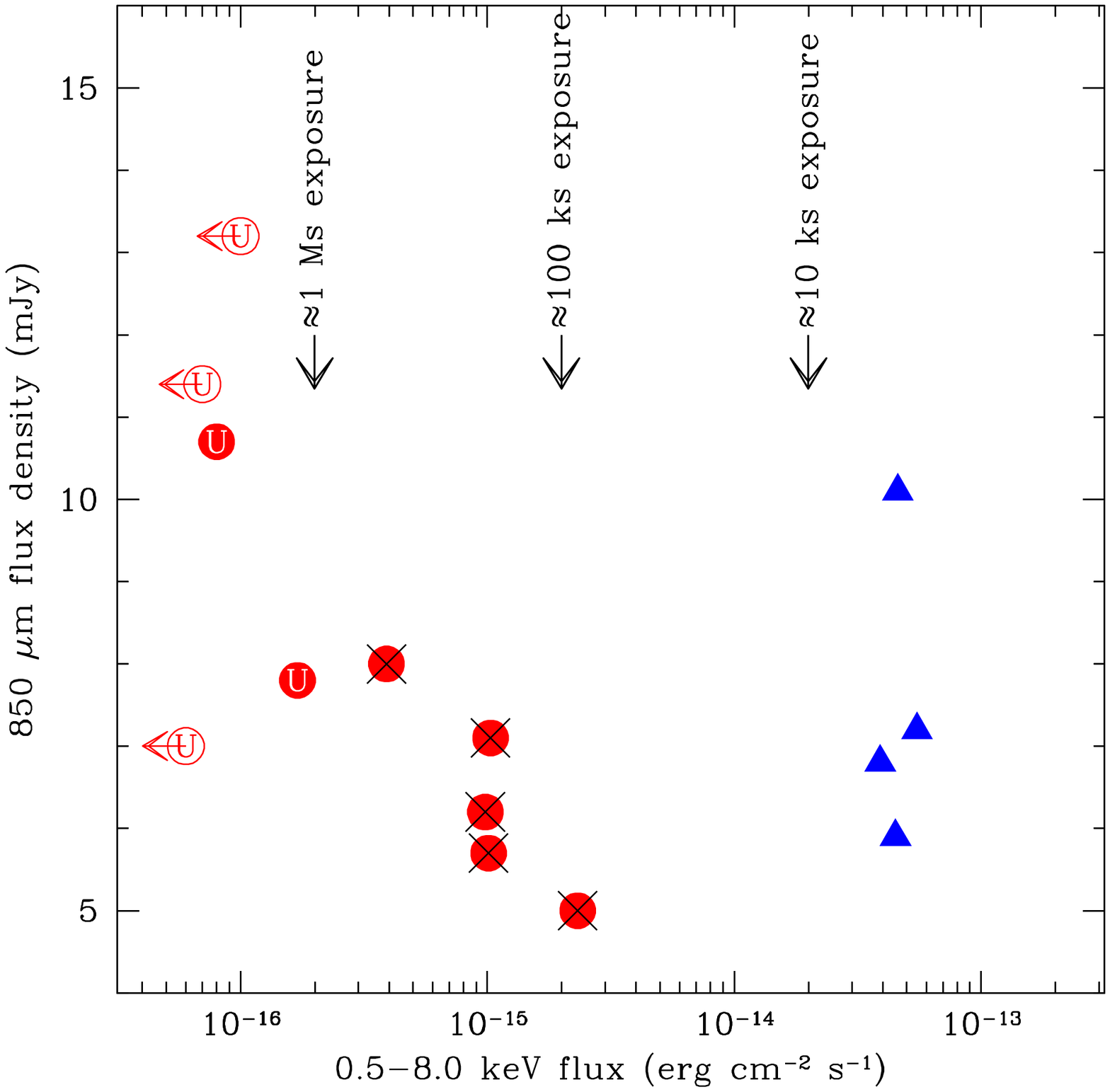}}
\vspace{0.1in}
\figcaption{Submm flux density versus full-band X-ray flux. The filled circles are the X-ray detected submm sources, and the open circles are the X-ray undetected submm sources (see \S2.4); crosses indicate sources classified as AGNs, and ``U'' indicates sources that do not have good X-ray spectral constraints (see \S3.2 and Table~2). The filled triangles are the submm-detected QSOs of Page \etal (2001); see \S3.1. The upper limit arrows in the $y$-axis direction show the approximate full-band flux limits for \chandra\ ACIS-I surveys of different exposure times. While submm-detected QSOs are detectable in $\simlt10$~ks \chandra\ exposures, $\simgt$100~ks \chandra\ exposures are required to detect typical submm sources.}
\label{fig:redshift1}
\vspace{0.2in}


Figure~3 shows evidence for an anti-correlation between the full-band flux and the submm flux density for our sources. While this anti-correlation could be real, we caution that it may be an artifact caused by small number statistics and sample incompleteness.

\subsection{X-ray Source Classification}

\subsubsection{X-ray Spectral Slopes and Luminosities}

The effective X-ray photon index ($\Gamma$) of a source can provide a simple constraint on its nature.\footnote{The photon index is related to the energy index by $\Gamma$~=~$\alpha$+1 where $F_{\nu}\propto\nu^{-\alpha}$.} Within the context of AGN-classified sources, obscured AGNs have considerably flatter effective X-ray spectral slopes than the canonical $\Gamma\approx2.0$ photon index of unobscured AGNs (e.g.,\ George \etal 2000), due to the energy-dependent photo-electric absorption of the X-ray emission (e.g.,\ Risaliti, Maiolino, \& Salvati 1999). Star-forming galaxies are also distinguishable from obscured AGNs since their rest frame $\approx$~2--8~keV emission is often consistent with that of a $\Gamma\approx2.0$ power law (e.g.,\ Kim, Fabbiano, \& Trinchieri 1992a; Ptak \etal 1999).\footnote{Detailed X-ray studies of local star-forming galaxies have shown that they have more complex X-ray spectra than simple power-law emission (e.g.,\ power-law emission and a very soft $\approx$~0.7~keV thermal component; Ptak \etal 1999); however, at the probable redshifts of our sources the very soft thermal component will be redshifted out of the \chandra\ energy band.}

The X-ray luminosity of a source can provide a further check on the nature of the X-ray emission since very few starburst galaxies have full-band luminosities in excess of \hbox{$L_{\rm X}\approx 10^{42}$~erg~s$^{-1}$}, even when including luminous sources at moderate redshifts (e.g.,\ Moran, Lehnert, \& Helfand
%
%
\centerline{\includegraphics[angle=0,width=9.0cm]{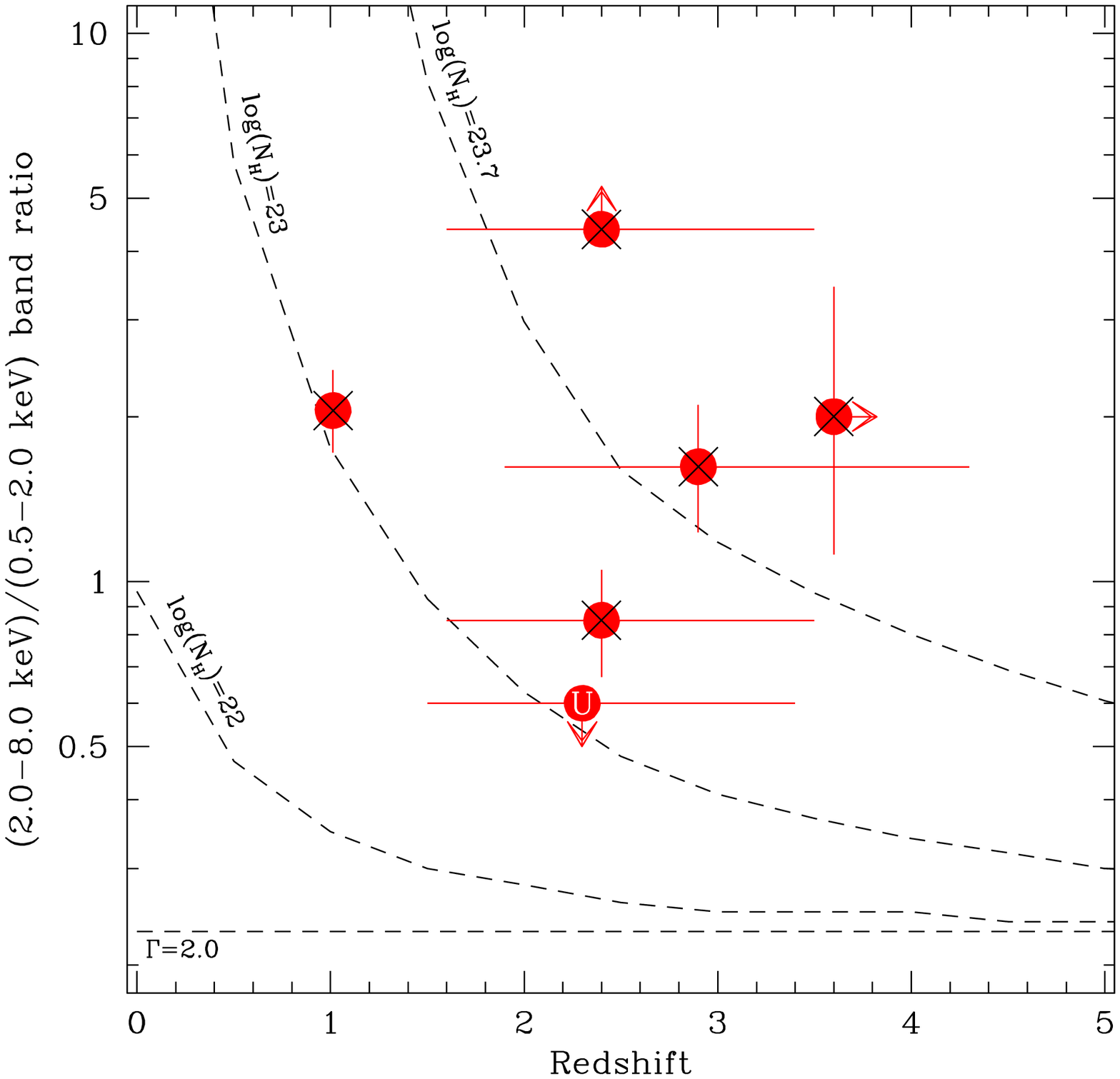}}
\vspace{0.02in}
\figcaption{X-ray band ratio versus redshift. The symbols have the same meaning as in Figure~3; the X-ray detected submm source 123618.3+621551 cannot be plotted here since it is only detected in the full band. The error bars in the $x$-axis direction correspond to the uncertainties in the millimetric redshifts, and the error bars in the $y$-axis direction correspond to the uncertainties in the band ratio. The dashed lines show the expected band ratios for different absorbing column densities with an underlying $\Gamma=2.0$ power-law continuum; these have been calculated using {\sc pimms} Version 3.2d. The AGN-classified sources are obscured, with implied intrinsic column densities of $N_{\rm H}\simgt 10^{23}$~cm$^{-2}$.}
\label{fig:redshift1}
\vspace{0.2in}

\noindent 1999; Zezas, Alonso-Herrero, \& Ward 2001; Alexander \etal 2002a). While this would usually be a reasonable upper limit to the expected X-ray emission from star-formation activity, given the extreme infrared luminosities of submm sources (\hbox{$L_{\rm IR}=10^{12}$--$10^{13}$~\Lsolar;} e.g.,\ Ivison \etal 1998; Smail \etal 2002), we might expect full-band luminosities up to an order of magnitude larger (i.e.,\ up to \hbox{$L_{\rm X}\approx 10^{43}$~erg~s$^{-1}$}) for this source class.\footnote{While $L_{\rm IR}=10^{12}$--$10^{13}$~\Lsolar\ galaxies are probably the best candidates for detecting $L_{\rm X}\approx 10^{43}$~erg~s$^{-1}$ starburst galaxies, none have been found to date.}

The X-ray band ratio versus redshift for the six sources with band ratio constraints is shown in Figure~4. Five of the X-ray detected submm sources have effective photon indices of $\Gamma<1.0$. Such flat X-ray spectral slopes are only likely to be produced by obscured AGN activity and, under the assumption that the underlying continuum follows a $\Gamma=2.0$ power law, these sources have intrinsic column densities of $N_{\rm H}\simgt$~$10^{23}$~cm$^{-2}$.

The rest-frame X-ray luminosities of these sources are calculated as

\begin{equation}
L_{\rm X}~=~4~\pi~d_{\rm L}^2~f_{\rm X}~(1+z)^{\Gamma-2}~\rm{erg~s^{-1}},
\end{equation}

\noindent where $d_{\rm L}$ is the luminosity distance (cm), $f_{\rm X}$ is the X-ray flux (erg~cm$^{-2}$~s$^{-1}$), and $\Gamma=2.0$ is the assumed photon index.

Assuming the underlying continuum follows a $\Gamma=2.0$ power law, the unabsorbed rest-frame full-band luminosities (\hbox{$L_{\rm X}\approx 10^{43}$--$10^{44}$~erg~s$^{-1}$}) of the five sources with flat X-ray spectral slopes are also consistent with obscured AGN activity (the intrinsic absorption is calculated from the band ratios; i.e.,\ see Figure~4). These sources are classified hereafter as AGNs (see Table~2).

The X-ray spectral constraints are weak for the other five submm sources. The rest-frame full-band luminosities or upper limits (\hbox{$L_{\rm X}\approx 10^{42}$--$10^{43}$~erg~s$^{-1}$}) for the three sources with luminosity constraints (see Table~2) suggest that their X-ray emission could be produced by luminous star-formation activity. However, since we cannot rule out AGNs in any of these five sources (e.g.,\ they could be extremely highly obscured or of low luminosity), they are classified here as unknown (see Table~2).

\subsubsection{X-ray Emission from Star Formation}

From a comparison of the X-ray and radio luminosities of starburst galaxies detected in the 1~Ms \chandra\ exposure of the CDF-N, Bauer \etal (2002b) showed that the radio emission can be used to predict the X-ray emission from star formation (see also Ranalli, Comastri, \& Setti 2002). The main assumption in this prediction is that the radio emission is dominated by star formation; the presence of an AGN component to the radio emission will lead to an overprediction of the contribution from star formation at X-ray energies.

None of the radio-detected submm sources is classified by Richards (1999) as an AGN on the basis of its radio emission although four sources are classified as unknown (see Table~1). Since only two of these four sources are classified as AGNs at X-ray energies, and all but one have steep radio spectral slopes, the radio emission is likely to be dominated by star-formation activity in most cases (e.g.,\ Condon 1992; Richards 2000). The rest-frame radio luminosity densities of these sources are calculated as

\begin{equation}
L_{\rm 1.4~GHz}~=~4~\pi~d_{\rm L}^2~f_{\rm 1.4~GHz}~10^{-36}~(1+z)^{\alpha-1}~\rm{W~Hz^{-1}},
\end{equation}

\noindent where $d_{\rm L}$ is the luminosity distance (cm), $f_{\rm 1.4~GHz}$ is the 1.4~GHz flux density ($\mu$Jy), and $\alpha$ corresponds to the radio spectral index. In determining the radio luminosity density we have assumed $\alpha=0.8$, the average spectral index for star-forming galaxies (e.g.,\ Yun, Reddy, \& Condon 2001).

A comparison of the rest-frame full-band X-ray and radio luminosities of the radio-detected submm sources is shown in Figure~5. The predicted X-ray luminosity from star-formation activity for all of the radio-detected submm sources is $L_{\rm X}\approx 10^{42}$--$10^{43}$~erg~s$^{-1}$. The AGN-classified sources are on average $\approx$~10 times more luminous at X-ray energies than that predicted by star formation. However, the X-ray emission from the other three radio-detected submm sources could be completely dominated by star-formation activity.

\subsection{X-ray Spectral Analysis}

Our interpretation of the X-ray spectral slopes and luminosities for the five AGN-classified sources (see \S3.2.1) was based on the assumption that the obscuration to the X-ray emitting source is not optically thick to Compton scattering. However, in some cases the obscuration may be Compton thick (i.e.,\ $N_{\rm H}>1.5\times10^{24}$~cm$^{-2}$), and the observed X-ray emission will be predominantly produced by reflection and scattering processes (e.g.,\ Matt \etal 1997; Turner \etal 1997a; Matt
%
%
\centerline{\includegraphics[angle=0,width=9.0cm]{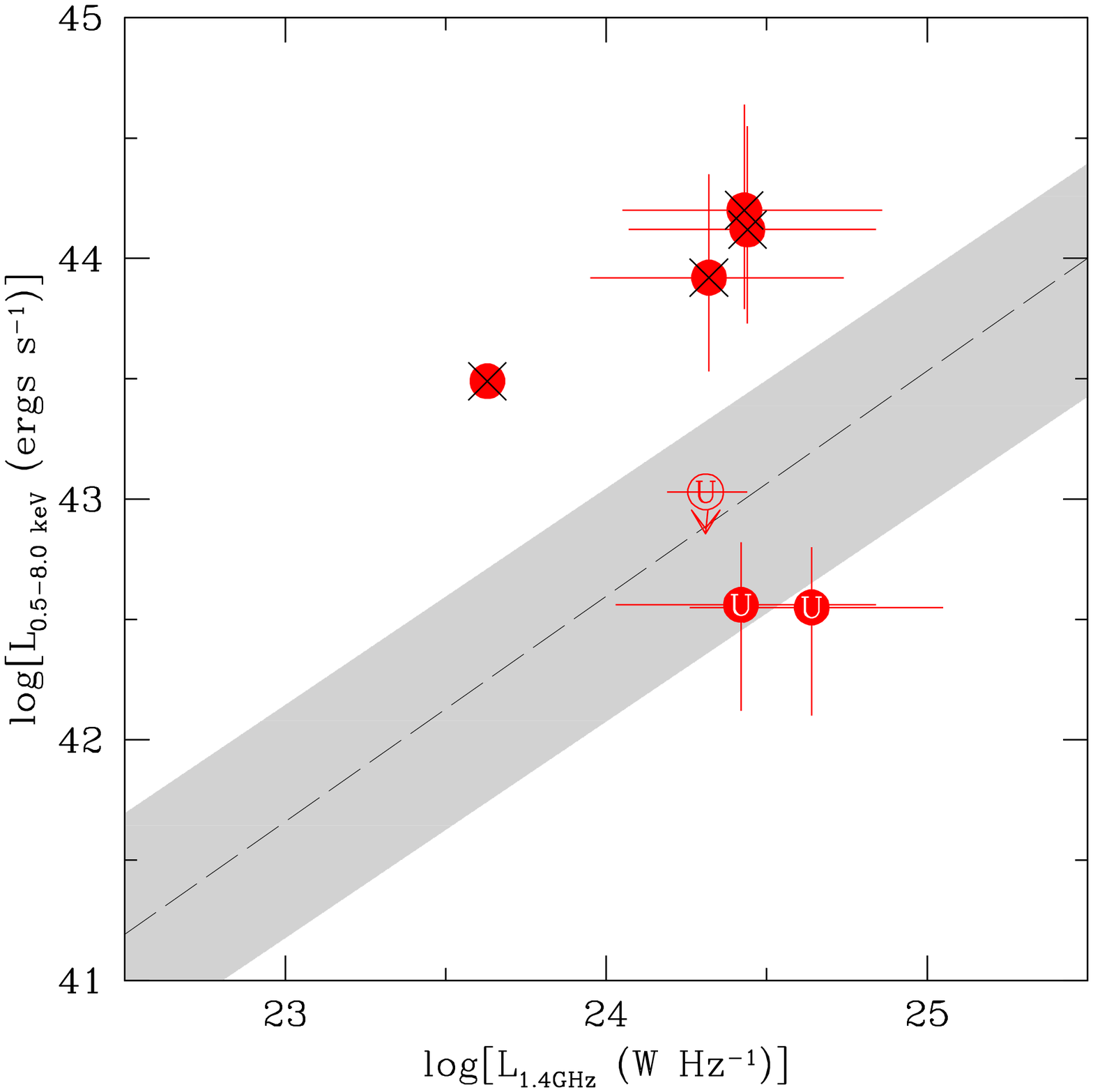}}
\vspace{0.1in}
\figcaption{Rest-frame full-band X-ray luminosity versus rest-frame 1.4~GHz radio luminosity density for the radio-detected submm sources. The radio-undetected sources cannot be plotted as they all have redshift lower limits. The symbols have the same meaning as in Figure~3. The error bars correspond to the uncertainties in the luminosities. The X-ray luminosities of the AGN-classified sources correspond to the unabsorbed luminosity. The shaded region denotes the 1~$\sigma$ dispersion in the locally-determined X-ray/radio correlation (see Figure~6 of Shapley, Fabbiano, \& Eskridge 2001). The dashed line is the regression fit to emission-line galaxies investigated in Bauer \etal (2002b) with the 1~Ms \chandra\ exposure of the CDF-N. The AGN-classified sources have X-ray luminosities on average $\approx$~10 times higher than those expected from star formation. The X-ray luminosities of the sources without good X-ray spectral constraints are consistent with those expected from star formation.}
\label{fig:redshift1}
\vspace{0.2in}


\noindent \etal 2000). The most direct discrimination between Compton-thin and Compton-thick absorption is made with X-ray spectral analysis. The X-ray spectrum of a Compton-thick AGN is generally characterized by a large equivalent width iron K$\alpha$ emission line ($EW\simgt$~0.5~keV; e.g.,\ Matt, Brandt, \& Fabian 1996; Maiolino \etal 1998; Bassani \etal 1999; Matt \etal 2000) and a flat or inverted ($\Gamma<0.0$) X-ray spectral slope, due to pure reflection. By contrast, the X-ray spectrum of a Compton-thin AGN is usually well fitted by an absorbed power-law model and a smaller equivalent width iron K$\alpha$ emission line (generally $EW\approx$~0.1--0.5~keV; Nandra \etal 1997; Turner \etal 1997b; Bassani \etal 1999).

To account for the range of roll angles and aim points in the 20 separate observations that comprise the 2~Ms CDF-N, we used the ACIS source extraction code ({\sc ACIS Extract}) described in Broos \etal (2002).\footnote{{\sc ACIS Extract} is a part of the {\sc TARA} software package and can be accessed from http://www.astro.psu.edu/xray/docs/TARA/ae\_users\_guide.html.} Briefly, for each source this code extracts the counts from each of the observations taking into account the changing shape and size of the PSF with off-axis angle as given in the \chandra\ X-ray Center (CXC) PSF library.\footnote{See http://asc.harvard.edu/ciao2.2/documents\_dictionary.html\#psf.} A local background is extracted after all sources are excluded from the X-ray event file, and the spectra and response matrices are summed using standard {\sc ftools} routines (Blackburn 1995). The latest version of {\sc xspec} (v11.2.0i; Arnaud 1996, 2002) was used for all of the model fitting, and the fit parameter uncertainties are quoted at the 90\% confidence level (i.e.,\ Avni 1976). Given the limited counting statistics for the AGN-classified sources ($\approx$~30--200 full-band counts), we performed all the X-ray spectral analysis using the $C$-statistic (Cash 1979). The advantage of using the $C$-statistic is that the data can be fitted unbinned, making it well suited to low-count sources (e.g.,\ Nousek \& Shue 1989). In the latest version of {\sc xspec}, the $C$-statistic can be used on background-subtracted data. We carried out several checks of the background-subtraction method to verify that no spurious residual features were present in the background-subtracted data.

We only considered two simple models for the X-ray spectral fitting: (1) a power-law model with Galactic absorption, and (2) a power-law with both Galactic and additional absorption. The results are given in Table~3. As expected, the photon indices derived using the Galactic absorbed power-law emission model are in good agreement (within $\pm0.2$) with those calculated from the band ratio. When fit with an absorbed power-law model, the best-fitting photon indices and observed absorption for three of the sources are in broad agreement with those expected for Compton-thin AGNs (e.g.,\ Turner \etal 1997b; Bassani \etal 1999), while two sources have particularly flat photon indices (123622.6+621629 and 123712.0+621211), possibly indicative of reflection and Compton-thick absorption. However, with only $\approx$~30 full-band counts, 123712.0+621211 is probably too faint to provide reliable constraints.

To investigate whether any of these sources could be Compton-thick AGNs we searched for emission-line features in the 2~$\sigma$ binned X-ray spectra. The only source that showed evidence for an emission-line feature was 123622.6+621629 (see Figure~6); this was also one of the sources with a flat photon index. To assess whether this $\approx$~4.4~keV emission-line feature was real, we simulated 10,000 low-count ($\approx$~90 full-band counts) X-ray spectra using the {\it fakeit} routine within {\sc xspec}; the input model was a Galactic absorbed $\Gamma=$~--0.7 power law. Thirty-four (0.34\%) of the simulations produced a feature with an equivalent width as large as or larger than that measured in the data, corresponding to a significance level of 2.9~$\sigma$. While this provides an estimation of the significance of this feature if we were expecting it to appear at $\approx$~4.4~keV, it does not take into account of the fact that we simply identified the most significant feature over the observed energy range of $\approx$~1.0--6.4~keV. To take this into account we must determine the number of possible distinguishable features over this energy range. Given the average spectral resolution of ACIS-I ($\approx$~150~eV) we would expect $\approx$~30 distinguishable features. However, by adding unresolved gaussian lines we estimate that we would only be able to distinguish six features in the X-ray spectrum of 123622.6+621629; this demonstrates that photon statistics are more important than spectral resolution in low-count sources. Thus, the adjusted significance level for this feature is $\approx$~2.3~$\sigma$ (i.e.,\ $6\times0.34$\%).

\subsubsection{123622.6+621629: a Compton-thick AGN?}

Assuming that the $\approx$~2.3~$\sigma$ emission feature is a redshifted, unresolved 6.4~keV iron K$\alpha$ emission line, the observed energy ($4.4\pm0.1$~keV) implies $z=$~0.46$^{+0.03}_{-0.02}$ and a rest-frame equivalent width of $0.7^{+0.6}_{-0.5}$~keV. The moderately large equivalent width of this possible 6.4~keV iron K$\alpha$ emission line, and the inverted X-ray spectral slope ($\Gamma=$~--0.7; see Table~3), provide evidence that 123622.6+621629 may be a Compton-thick AGN. Under the assumption that the underlying AGN model is
%
%
\centerline{\includegraphics[angle=-90,width=9.0cm]{alexander.fig6.ps}}
\vspace{0.1in}
\figcaption{The X-ray spectrum of 123622.6+621629. The data are shown here in 2~$\sigma$ bins for presentation purposes; the error bars are Gaussian and underestimate the true errors by $\approx$~30\%. The top panel shows the X-ray spectrum with a fitted model of Galactic absorbed power-law emission ($\Gamma$~=--0.7; see Table~3). The bottom panel shows the ratio between the X-ray spectrum and the fitted model. Under the assumption that the $\approx$~2.3~$\sigma$ feature at $4.4\pm0.1$~keV is the 6.4~keV iron K$\alpha$ emission line, the inferred source redshift is $z=$~0.46$^{+0.03}_{-0.02}$ and the large rest-frame equivalent width ($0.7^{+0.6}_{-0.5}$~keV) provides evidence that 123622.6+621629 may be a Compton-thick AGN (see \S3.3.1).}
\label{fig:redshift1}
\vspace{0.2in}


\noindent the same as that fitted to the Compton-thick AGN NGC~6240 (Vignati \etal 1999), the unabsorbed rest-frame full-band luminosity at $z\approx$~0.46 is $L_{\rm X}\approx 10^{43}$~erg~s$^{-1}$.

The inferred redshift ($z=$~0.46$^{+0.03}_{-0.02}$) is in considerable disagreement with the millimetric redshift ($z=$~2.4$^{+1.1}_{-0.8}$). However, we note that since 123622.6+621629 is detected in both the $U$ band (Barger \etal 2000) and the $U_{\rm n}$ band (Hogg \etal 2000), it is actually likely to lie at $z<2$ (to avoid the Lyman break). With a faint optical counterpart ($I=23.4$), the absolute optical magnitude of 123622.6+621629 at $z\approx$~0.46 would correspond to that expected from a dwarf galaxy (i.e.,\ $\approx$~3--4 mags below $L_*$). There are no known examples of luminous AGNs in dwarf galaxies in the local universe (e.g.,\ McLeod \& Rieke 1995), although it is possible that 123622.6+621629 is an $L_*$ galaxy with $\approx$~3--4 mags of dust extinction.

If 123622.6+621629 lies at redshift of $z\approx$~0.46 then its properties are atypical for a submm source (e.g.,\ Ivison \etal 2000; Smail \etal 2002). However, we note that 123622.6+621629 shares many similar characteristics with FN1-40, a $z=0.45$ submm source with an $I=24$ optical counterpart (Chapman \etal 2002a). Although good X-ray constraints do not exist for FN1-40, the lack of broad optical emission lines led Chapman \etal (2002a) to claim no evidence for AGN activity. While FN1-40 is clearly not a broad-line AGN, we note that with an [OIII]~$\lambda$5007/H$\beta$ emission line ratio of $\simgt$~3 (see Figure 2 of Chapman \etal 2002a), it could be a narrow-line AGN (e.g.,\ Baldwin, Phillips, \& Terlevich 1981; Veilleux \& Osterbrock 1987). Hence, FN1-40 and 123622.6+621629 may be similar objects. Optical spectroscopy is clearly required to determine the redshift of 123622.6+621629 and help shed light on the origin of its $\approx$~2.3~$\sigma$ emission line feature.

%
\section{The Origin of the Submm Emission}\label{origin}
%

%
%
\begin{figure*}[t]
\vspace{0.2in}
\centerline{\includegraphics[angle=0,width=11.0cm]{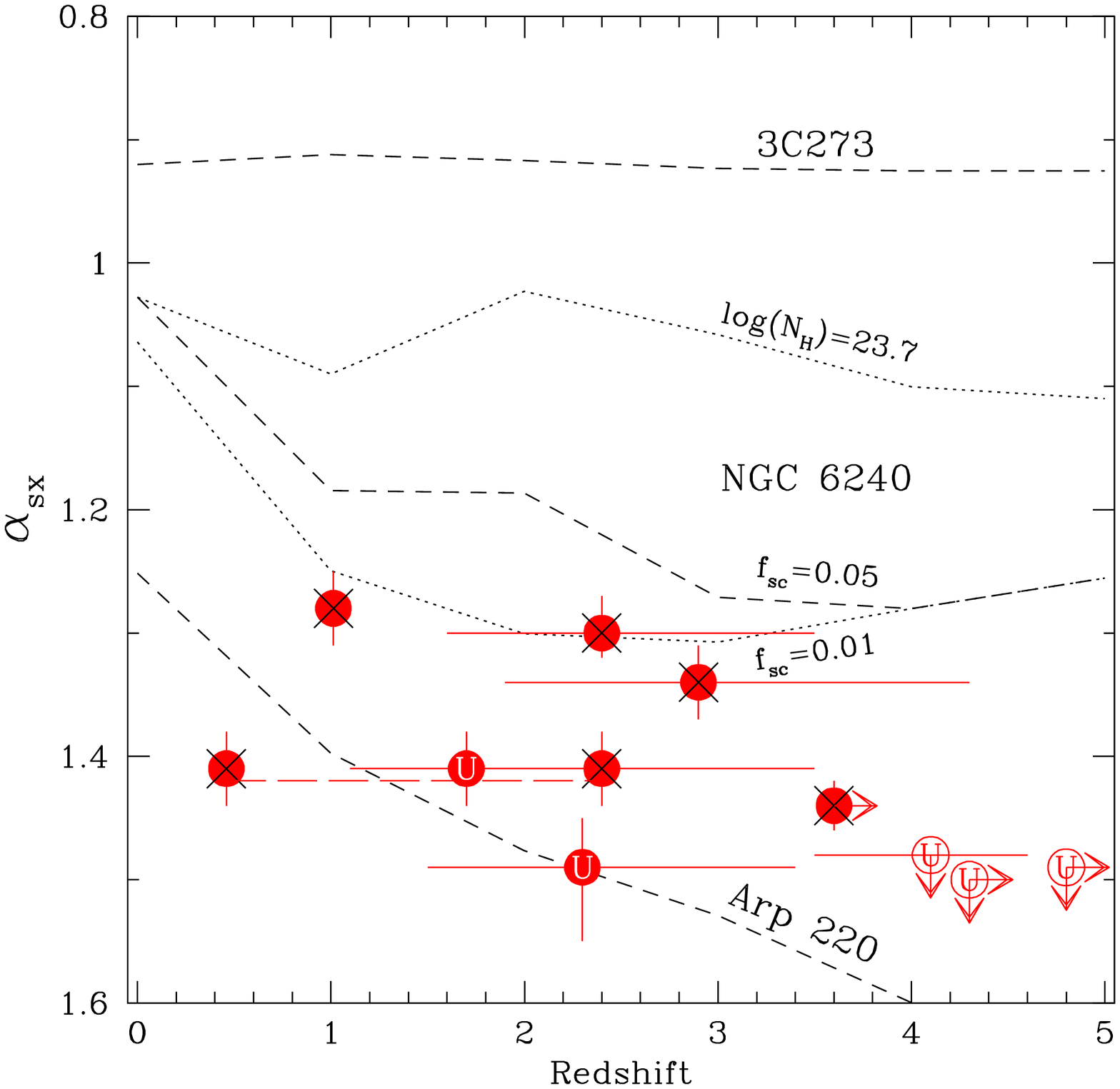}}
\vspace{0.1in}
\figcaption{Submm-to-X-ray spectral index ($\alpha_{\rm SX}$) versus redshift. The symbols have the same meaning as in Figure~3; the long-dashed line connecting two filled circles indicates the two possible redshifts for 123622.6+621629 (i.e.,\ $z=0.46^{+0.03}_{-0.02}$ and $z=2.4^{+1.1}_{-0.8}$). The error bars in the $x$-axis direction correspond to the uncertainties in the millimetric redshifts, and the error bars in the $y$-axis direction correspond to the uncertainties in $\alpha_{\rm SX}$. The X-ray flux density corresponds to 2~keV and is calculated from the full-band flux using the measured photon index of each source. Overplotted are the expected $\alpha_{\rm SX}$ values for 3C273, NGC~6240 ($f_{sc}=$~0.05), and Arp~220 (dashed lines); alternative curves for NGC~6240 with less internal absorption ($N_{\rm H}=5\times10^{23}$~cm$^{2}$) and a smaller scattered flux fraction ($f_{sc}=$~0.01) are also shown (dotted lines). This figure has been adapted from Fabian \etal (2000); the references for the 3C273, NGC~6240 and Arp~220 data are Neugebauer, Soifer, \& Miley (1985), Iwasawa (1999), Vignati \etal (1999), and Sanders (2000). Although five of the X-ray detected submm sources are classified as AGNs, their submm emission is likely to be dominated by star formation.}
\label{fig:redshift1}
\end{figure*}


It is often assumed that the submm emission from sources detected in \scuba\ surveys can provide an unobscured indicator of the star-formation rate (e.g.,\ Blain \etal 1999). However, five of the ten submm sources investigated here clearly host obscured AGNs. While the AGNs obviously dominate the X-ray emission in these sources, this does not necessarily imply that the submm emission is also powered by the AGNs. Unfortunately, the broad-band submm emission itself cannot provide an unambiguous diagnostic of the nature of the dominant physical mechanism since it is likely to be reprocessed thermal emission produced via the absorption of the primary emission by dust grains.\footnote{Further insight into the origin of the submm emission may come from determining its spatial distribution (i.e.,\ AGN emission is typically confined to the central galactic regions, while star formation can be more widespread). Such observations should be possible with the advent of the Atacama Large Millimeter Array (ALMA; see http://www.alma.nrao.edu/) in Chile and the Sub-Millimeter Array (SMA; see http://sma-www.harvard.edu/) in Hawaii.}

CO observations can provide some insight into the origin of the submm emission by measuring the molecular gas mass and hence constrain the expected dust mass and emission from star formation (e.g.,\ Young \& Scoville 1991). For instance, CO observations have been used to show that star-formation activity may account for $\approx$~50\% of the submm emission from the $z=2.80$ BALQSO SMM J02399--0136 (e.g.,\ Frayer \etal 1998) and perhaps all of the submm emission from the $z=2.56$ galaxy SMM J14011+0252 (e.g.,\ Frayer \etal 1999; Ivison \etal 2001). However, in the absence of CO observations we must adopt an indirect strategy in estimating the origin of the submm emission from our submm sources.

\subsection{Constraints from the Submm-to-X-ray Spectral Slopes}

The relative importance of AGN and star-formation activity in our submm sources can be estimated by comparing their submm-to-X-ray spectral slopes ($\alpha_{\rm SX}$) to those of well-studied nearby galaxies. We calculate $\alpha_{\rm SX}$ as

\begin{equation}
\alpha_{\rm SX}~=~-~[\log({f_{\rm 2keV}\over{f_{\rm 850\mu m}}})~-~0.18]~\times~0.163,
\end{equation}

\noindent where $f_{\rm 2keV}$ is the observed flux density at 2~keV (keV~cm$^{-2}$~s$^{-1}$~keV$^{-1}$), calculated from the full-band flux using the estimated photon index, and $f_{\rm 850\mu m}$ is the observed flux density at 850~$\mu$m (mJy). Larger values of $\alpha_{\rm SX}$ indicate stronger submm emission relative to the X-ray emission.

Following Fabian \etal (2000) we have compared our sources to the QSO 3C273, the Compton-thick AGN NGC~6240, and the archetypal ultra-luminous starburst galaxy Arp~220. While 3C273 is clearly a powerful unobscured QSO, NGC~6240 contains components of obscured AGN and star-formation activity. By contrast, the emission from Arp~220 appears to be entirely dominated by star formation. If an AGN is present in Arp~220, then to remain undetected at X-ray energies it must be either of low luminosity, or have extreme absorption (i.e.,\ $N_{\rm H}\approx10^{25}$--$10^{26}$~cm$^{-2}$) and a small component of scattered emission (e.g.,\ Iwasawa \etal 2001, 2002; Clements \etal 2002). Since NGC~6240 has a considerably flatter X-ray spectral slope than 3C273 and Arp~220, the analysis of the submm-to-X-ray spectral slopes requires knowledge of the X-ray photon index for each source. For instance, at a given submm luminosity, the 2~keV flux density of a $\Gamma=0.0$ source will be $\approx$~5 times lower than that of a $\Gamma=2.0$ source. Therefore, when interpreting the data for an unobscured QSO, the comparison should be made to 3C273, while when interpreting the data for an obscured AGN, the comparison should be made to NGC~6240.

The submm-to-X-ray spectral slope versus redshift of all the sources is shown in Figure~7. The AGN-classified sources have larger $\alpha_{\rm SX}$ values than NGC~6240, suggesting that their AGNs are comparitively weaker at X-ray energies (with respect to their submm emission). Under the assumption that the obscuration in these sources is Compton thin (as expected for the majority of these sources; see \S3.3), they are $\approx$~15--150 times less luminous at X-ray energies (for a given submm luminosity) than NGC~6240 (assuming a column density of $N_{\rm H}=5\times10^{23}$~cm$^{-2}$; i.e.,\ the top dotted line in Figure~7). Taking all of the AGN-classified sources together, they are on average $\approx$~70 times less luminous at X-ray energies (for a given submm luminosity) than NGC~6240. As can be seen from Figure~7, these results are relatively insensitive to redshift so long as the AGN-classified sources lie at $z>1$. The only source where we have evidence that this may not be the case is the possible Compton-thick AGN 123622.6+621629 (see \S3.3.1 and Figure~6). Under the assumption that the obscuration to this source is the same as that found for NGC~6240 (i.e.,\ Compton-thick absorption with a 5\% scattering fraction; Vignati \etal 1999), at $z\approx$~0.46 it would be $\approx$~80 times less luminous at X-ray energies (for a given submm luminosity) than NGC~6240 (see Figure~7). 

We suggested in \S3 that the X-ray emission from the submm sources without good X-ray constraints may be produced by star-formation activity in some cases. This hypothesis is further supported for the two sources with millimetric redshifts of $z<4$ since their submm-to-X-ray spectral slopes are comparable to that expected from Arp~220 at $z>1$ (see Figure~7). However, we cannot distinguish between star-formation and AGN scenarios for the three $z>4$ sources.

\subsection{Constraints from NGC~6240}

Although the individual constraints on the origin of the submm emission from the AGN-classified sources are weak, they all clearly have larger submm-to-X-ray spectral slopes than NGC~6240. Even though NGC~6240 is a nearby source ($\approx$~115~Mpc in our chosen cosmology), the origin of its submm emission is poorly known. For instance, the application of radiative transfer models shows that the far-IR-to-submm emission can be almost completely explained by starburst activity (Rowan-Robinson \& Crawford 1989), and the infrared (imaging and spectroscopy) characteristics of NGC~6240 suggest that star formation dominates the bolometric luminosity (Genzel \etal 1998). Furthermore, its nuclear optical spectrum shows the characteristics of a LINER (e.g.,\ Veilleux \etal 1995). Conversely, consideration of the energetics of the AGN derived from X-ray observations suggests that it may contribute $\approx$~50\% of the far-IR-to-submm luminosity (e.g.,\ Iwasawa \& Comastri 1998; Vignati \etal 1999; Lira \etal 2002).

Assuming that the AGN completely dominates the submm emission in NGC~6240 (an extreme hypothesis) and the X-ray emission from the AGN scales linearly with submm luminosity, the contribution to the submm emission from AGN activity in the AGN-classified sources would be 0.6--6.6\% (as determined from the $\approx$~15--150 times difference in X-ray luminosity for a given submm luminosity; see \S4.1). Following the same assumptions, the average contribution to the submm emission from AGN activity would be $\approx$~1.4\%. Hence, although there is considerable uncertainty in the AGN contribution at submm wavelengths, these analyses strongly suggest that the submm emission is dominated by star-formation activity. This conclusion is likely to be valid even if the $z>4$ sources are also found to contain AGNs (see Figure~7).

%
\section{Discussion}\label{disc}
%

It has been known for some time that AGNs comprise a fraction of the submm source population. Indeed, the first identified source in a \scuba\ blank-field survey was found to be an AGN (SMM J02399--0136; Ivison \etal 1998), and at least three of the 15 submm-detected sources in the \scuba-lens survey are AGNs (Smail \etal 2002). In all these cases the evidence for AGN activity was found via optical spectroscopy of optically bright sources. The advantage of searching for AGNs with X-ray observations is that the limitations of optical spectroscopy are removed, allowing optically faint AGNs and AGNs without strong emission lines to be identified (e.g.,\ Vignati \etal 1999; Mushotzky \etal 2000; Alexander \etal 2001; Hornschemeier \etal 2001). However, obviously the lack of optical spectroscopy for the majority of our sources leads to considerable uncertainties in their source redshifts. In this final section, we discuss the dependence of our results on the assumed source redshifts, the fraction and properties of AGNs in the submm source population, and the AGN contribution to the submm emission in \scuba\ sources.

\subsection{Dependence of the Results on Source Redshifts}

The redshifts for the majority of the submm sources are derived from the radio-to-submm spectral slope and are uncertain. As mentioned in \S4.1 these uncertainties should not change our main result (i.e.,\ that the average AGN contribution to the submm emission is of the order of $\approx$~1.4\%) so long as the sources lie at $z>1$. However, we note here that since the redshift tracks for the template galaxies rise to lower values of $\alpha_{\rm SX}$ at $z<1$ (see Figure~7), the case for star formation dominating the submm emission is even stronger for sources at lower redshifts (as may be the case for \hbox{123622.6+621629}; see \S3.3.1). Finally, we also note that although the implied column densities and X-ray luminosities of the AGN-classified sources will be lower if they lie at $z<1$, since $\Gamma<1.0$ power-law emission is only likely to be produced by obscured AGN activity, their classification as AGNs is insensitive to redshift.

\subsection{AGNs in the Bright Submm Source Population}

In this study we have shown that five of the seven X-ray detected submm sources appear to be obscured AGNs, placing a lower limit on the source density of obscured AGNs with $f_{\rm 850\mu m}\ge$~5~mJy of $260^{+170}_{-110}$~deg$^{-2}$. Under the same assumptions as those used in determining the fraction of the bright submm source population with X-ray counterparts (see \S2.4.2), this suggests that $\simgt$~26\% of the bright submm source population host AGNs (in reasonable agreement with model predictions; e.g.,\ Almaini, Lawrence, \& Boyle 1999; Fabian \& Iwasawa 1999).

Ivison \etal (2002) found a similar AGN fraction to that derived here using the moderately deep \xmm\ observations of the Lockman Hole East region (e.g.,\ Hasinger \etal 2001). However, their submm-detected AGN source density is $\approx$~2 times smaller (i.e.,\ $120^{+90}_{-50}$~deg$^{-2}$ versus $260^{+170}_{-110}$~deg$^{-2}$).\footnote{Bautz \etal (2000) also found a similar AGN fraction from combining their two \chandra\ detected submm sources with the \chandra\ undetected submm sources in the HDF-N (e.g.,\ Hornschemeier \etal 2000), although their AGN fraction may be biased by the inclusion of a comparitively rare submm-detected QSO (the BALQSO SMM J02399--0136; Vernet \& Cimatti 2001).} The difference in the submm-detected AGN source densities is mostly due to the higher sensitivity of our X-ray observations. Hence the similarity between the AGN fraction found by Ivison \etal (2002) and our lower limit may indicate that our estimated AGN fraction is conservative.

The unabsorbed rest-frame full-band luminosities of the AGN-classified sources are $L_{\rm X}\approx 10^{43}$--$10^{44}$~erg~s$^{-1}$. This range of X-ray luminosity is typical of luminous Seyfert galaxies. Therefore, although the majority of the submm sources are likely to be extremely luminous infrared galaxies (e.g.,\ Ivison \etal 1998; Smail \etal 2002), their AGNs are comparitively weak. For example, typical QSOs of similar infrared luminosity (e.g.,\ Isaak \etal 2002; Page \etal 2001) are $\approx$~10--100 times more luminous at X-ray energies (e.g.,\  Page \etal 2001; Vignali \etal 2001, 2003). Under the assumption that bright submm sources are massive proto galaxies (e.g.,\ Blain \etal 1999; Lilly \etal 1999), the luminosities of these AGNs are consistent with those expected for growing supermassive black holes (e.g.,\ Archibald \etal 2002).

\subsection{AGN Contribution to the Submm Emission}

The current constraints suggest that the total contribution to the submm emission from AGN activity is unlikely to be more than $\approx$~1.4\%; however, as noted in \S2.2, there are likely to be further bright submm sources in our region that have not yet been detected by \scuba. In addition, we have not considered the possible contribution from optically bright submm-detected QSOs. Although optically bright submm-detected QSOs may comprise only $\approx$~5--10\% of the bright submm source population (see footnote 1), their submm emission could have a large AGN component (perhaps of the order of $\approx$~50\%; Frayer \etal 1998; Bautz \etal 2000). Hence, although optically bright submm-detected QSOs may not be numerically significant, they may provide the bulk of the AGN emission at submm wavelengths.

%
\section{Conclusions}\label{con}
%

We have used the 2~Ms \chandra\ exposure of the CDF-N to constrain the X-ray properties of bright \scuba\ sources ($f_{\rm 850\mu m}\ge$~5~mJy; S/N$\ge4$). In this study we have focused on the X-ray spectral properties of the X-ray detected submm sources to determine whether the AGN-classified sources are Compton thick or Compton thin. We have used these results to constrain the contribution to the submm emission from AGN activity. Our main results are the following:

\begin{itemize}

\item Seven of the ten bright submm sources are detected with X-ray emission. The corresponding source density of bright submm sources with X-ray counterparts ($360^{+190}_{-130}$~deg$^{-2}$) suggests that $\simgt$~36\% of bright submm sources have X-ray counterparts at this X-ray depth; however we note that this fraction is somewhat uncertain since this region does not have complete \scuba\ coverage down to $f_{\rm 850\mu m}=$~5~mJy. See \S2.

\item Five of the X-ray detected submm sources have flat X-ray spectral slopes ($\Gamma<1.0$) and luminous X-ray emission ($L_{\rm X}\approx 10^{43}$--$10^{44}$~erg~s$^{-1}$), suggesting obscured AGN activity with $N_{\rm H}\simgt10^{23}$~cm$^{-2}$. One source is possibly a Compton-thick AGN since it has an extremely flat X-ray spectral slope ($\Gamma\approx$~--0.7) and shows possible evidence ($\approx$~2.3~$\sigma$) for a redshifted 6.4~keV iron K$\alpha$ emission line; however, the inferred redshift ($z=$~0.46$^{+0.03}_{-0.02}$) is in considerable disagreement with the millimetric redshift ($z=2.4^{+1.1}_{-0.8}$). See \S3.

\item A comparison of the five AGN-classified sources to the well-studied heavily obscured AGN NGC~6240 suggests that the AGNs in these sources contribute, on average, a negligible fraction (i.e.,\ $\approx$~1.4\%) of the submm emission. Hence, the submm emission from these sources appears to be dominated by star formation. This result is relatively insensitive to redshift. See \S4.

\item The X-ray constraints are weak for the other five submm sources. We find that the X-ray properties for the two sources with millimetric redshifts of $z<4$ are consistent with those expected from luminous star-formation activity ($L_{\rm X}\approx 10^{42}$--$10^{43}$~erg~s$^{-1}$). The submm-to-X-ray spectral slopes of these sources are also similar to those expected from Arp~220 (the archetypal ultra-luminous dusty starburst galaxy) at $z>1$; however, we cannot rule out the possibility that low-luminosity AGNs are present in these sources. The three $z>4$ sources could be either AGNs or starbust galaxies. See \S3 and \S4.

\item The fraction of the bright submm source population with AGN activity (i.e.,\ $\simgt$~26\%) is in reasonable agreement with model predictions. The comparitively low X-ray luminosities of the AGN-classified sources are more consistent with Seyfert galaxies than QSOs. We suggest that optically bright submm-detected QSOs (i.e.,\ luminous type 1 AGNs), although possibly not numerically significant, may provide the bulk of the AGN emission at submm wavelengths. See \S5.

\end{itemize}

These results show that the low X-ray detection rate of bright submm sources by moderately deep X-ray surveys is probably due to the relatively low luminosities of the AGNs in these sources rather than Compton-thick absorption. The current constraints therefore suggest that the total contribution from AGNs to the submm emission in the bright submm source population is negligible. Hence, the submm emission of bright submm sources appears to be predominantly powered by star-formation activity, and therefore can be used to determine star-formation rates.

%
\section*{Acknowledgments}
%

This work would not have been possible without the support of the
entire \chandra\ and ACIS teams; we particularly thank Pat Broos and
Leisa Townsley for data analysis software, CTI correction support, and
the development of {\sc ACIS Extract}. Special thanks go to Rob Ivison and
Ian Smail for detailed and insightful comments on various drafts of
this paper. We also thank Omar Almaini, Mike Eracleous, Konstantin
Getman, Giorgio Matt, Matt Page, Susie Scott, Paola Severgnini, Chris
Willott, and the attendees of the Durham-Edinburgh Extragalactic
Workshop on ``The First Galaxies and AGN'' for fruitful
discussions. We acknowledge the financial support of NASA grants
NAS~8-38252 and NAS~8-01128 (GPG, PI), NSF CAREER award AST-9983783
(DMA, FEB, WNB, CV), CXC grant G02-3187A (DMA, FEB, WNB, CV), NASA
GSRP grant NGT5-50247 (AEH), NSF grant AST-9900703~(DPS), NASA GSRP
grant NGT5-50277 (SCG), and the Pennsylvania Space Grant Consortium
(AEH, SCG).

%

%
%

\newpage

%
%

\begin{deluxetable}{llccccccccccc}
\rotate
\tablecolumns{13}
\tabletypesize{\scriptsize}
\tablewidth{0pt}
\tablecaption{Basic properties of the bright submm sources}
\tablehead{
\multicolumn{2}{c}{Coordinates} &
\colhead{Radio} &
\colhead{X-ray} &
\colhead{} &
\colhead{} &
\colhead{Radio} &
\colhead{} &
\colhead{} &
\colhead{} &
\colhead{Submm} &
\multicolumn{2}{c}{Derived properties} \\
\colhead{$\alpha_{2000}$$^{\rm a}$}       &
\colhead{$\delta_{2000}$$^{\rm a}$}       &
\colhead{detected?}   &
\colhead{detected?}   &
\colhead{$S_{\rm 1.4~GHz}$$^{\rm b}$}&
\colhead{$\alpha$$^{\rm b}$}    &
\colhead{ID$^{\rm c}$}          &
\colhead{$I^{\rm d}$}           &
\colhead{$I-K^{\rm d}$}         &
\colhead{$S_{\rm 850\mu m}$$^{\rm e}$}&
\colhead{Ref$^{\rm f}$}&
\colhead{$z^{\rm g}$}           &
\colhead{$L_{\rm 1.4~GHz}$$^{\rm h}$}}
\startdata
12 36 16.15 & +62 15 13.7 & Y & Y & 54$\pm$8   & $-$          & S & 24.3    & $<$3.2 & 5.7$\pm1.1$ & B02 & 2.4$^{+1.1}_{-0.8}$ & $2.1^{+2.8}_{-1.3}$ \\

12 36 18.33 & +62 15 50.5 & Y & Y & 151$\pm$11 & $>$0.63      & U & $>$25.3 & $-$    & 7.8$\pm1.6$ & B00 & 1.7$^{+0.8}_{-0.6}$ & $2.7^{+3.8}_{-1.7}$ \\

12 36 20.3  & +62 17 01   & N & N & $<40$      & $-$          & $-$ & $-$   & $-$    & 13.2$\pm2.9$ & BY & $>$4.8          & $-$ \\

12 36 21.1  & +62 12 50   & N & N & $<40$      & $-$          & $-$ & $-$   & $-$    & 11.4$\pm2.8$ & BY & $>$4.3          & $-$ \\

12 36 22.65 & +62 16 29.7 & Y & Y & 71$\pm$9   & $-$          & S & 23.4    & $<$2.5 & 7.1$\pm1.7$ & B00 & 2.4$^{+1.1}_{-0.8}$ & $2.8^{+3.7}_{-1.7}$  \\

12 36 29.13 & +62 10 45.8 & Y & Y & 81$\pm$9   & $>$0.80      & U & 22.2    & 3.8    & 5.0$\pm1.2$ & B02 & 1.013     & 0.4 \\

12 36 46.05 & +62 14 48.7 & Y & Y & 124$\pm$10 & 0.84$\pm$0.12& U & 24.9    & $<$3.8 & 10.7$\pm2.1$& B02 & 2.3$^{+1.1}_{-0.8}$ & $4.4^{+6.1}_{-2.7}$  \\

12 36 52.06 & +62 12 25.7 & Y & N$^{\rm i}$ & 16$\pm$4   & 0.42$^{+0.32}_{-0.30}$ & $-$ & $>$28.6 & $>$5.2 & 7.0$\pm0.5$ & HU & 4.1$^{+0.5}_{-0.6}$ & $2.1^{+0.6}_{-0.6}$ \\

12 37 07.21 & +62 14 08.1 & Y & Y & 45$\pm$8   & 0.29$\pm$0.16& U & 25.0    & 5.0    & 6.2$\pm1.3$ & B02 & 2.9$^{+1.4}_{-1.0}$ & $2.7^{+3.7}_{-1.7}$  \\

12 37 12.07 & +62 12 11.3 & N & Y & $<40$      & $-$          & $-$ & $>$25.3  & $-$ & 8.0$\pm1.8$ & B02 & $>$3.6            & $-$ \\ 

\enddata

\tablenotetext{a}{Source coordinates. Taken from, in order of preference, the radio position (if radio detected), the X-ray position (if X-ray detected), or the submm source position.}
\tablenotetext{b}{Radio flux density at 1.4 GHz in units of $\mu$Jy and radio spectral index ($\alpha$). The radio flux density upper limits correspond to the 5~$\sigma$ survey limit given in Richards (2000). The radio spectral index is defined from the 8.5~GHz and 1.4~GHz flux densities as $F_{\nu}\propto\nu^{-\alpha}$. Taken from Richards (2000) and Dunlop \etal (2002).}
\tablenotetext{c}{Source classification based on radio properties for sources detected in Richards (1999). ``S'' indicates star-forming galaxy, and ``U'' indicates that good constraints could not be placed.}
\tablenotetext{d}{$I$-band Vega-based magnitude and $I-K$ color. The calculated $I-K$ color is determined from $I-(HK^\prime-0.3)$ following Barger et~al. (1999b) for all sources except 123652.0+621225. The $I$-band magnitude and $I-K$ color for 123652.0+621225 are taken from Dunlop et~al. (2002).}
\tablenotetext{e}{Submm flux density at 850~$\mu$m in units of mJy.}
\tablenotetext{f}{Reference for submm data (HU=Hughes et~al. 1998; B00=Barger et~al. 2000; B02=Barger et~al. 2002; BY=Borys et~al. 2002).}
\tablenotetext{g}{Source redshift, either spectroscopic (Cohen et~al. 2000; Hornschemeier et~al. 2001), based on the source multi-wavelength SED (Dunlop et~al. 2002), or millimetric (Carilli \& Yun 2000); only 123629.1+621045 has a spectroscopic redshift ($z=$~1.013) and only 123652.0+621225 has an SED-based redshift ($z=$~4.1$^{+0.5}_{-0.6}$). The uncertainties in the SED and millimetric redshifts correspond to the 1$\sigma$ level. We note that the source redshift of 123622.6+621629 may be $z\approx$~0.46, based on X-ray spectral analysis (see \S3.3.1).}
\tablenotetext{h}{Rest-frame 1.4~GHz radio luminosity density in units of $10^{24}$ W~Hz$^{-1}$ determined following equation 2. The uncertainties on the calculated luminosity densities are determined using the upper and lower redshift bounds. We are unable to place luminosity constraints for the three radio-undetected sources as they all have redshift lower limits.}
\tablenotetext{i}{This source may have been detected at low-significance in the 1~Ms CDF-N study of VROs (Alexander \etal 2002b); see \S2.4.}

\newpage

\end{deluxetable}

%
%

\begin{deluxetable}{llccccccccccc}
\rotate
\tablecolumns{13}
\tabletypesize{\scriptsize}
\tablewidth{0pt}
\tablecaption{X-ray properties of the bright submm sources}
\tablehead{
\multicolumn{2}{c}{Coordinates} &
\multicolumn{3}{c}{X-ray counts}      &
\colhead{Band} &
\colhead{Effective} &
\multicolumn{3}{c}{X-ray flux} &
\multicolumn{2}{c}{Derived properties} &
\colhead{X-ray}\\
\colhead{$\alpha_{2000}$$^{\rm a}$}       &
\colhead{$\delta_{2000}$$^{\rm a}$}       &
\colhead{FB$^{\rm b}$}          &
\colhead{SB$^{\rm b}$}          &
\colhead{HB$^{\rm b}$}          &
\colhead{Ratio$^{\rm c}$}       &
\colhead{$\Gamma$$^{\rm d}$}    &
\colhead{FB$^{\rm e}$}          &
\colhead{SB$^{\rm e}$}          &
\colhead{HB$^{\rm e}$}          &
\colhead{$\alpha_{\rm SX}$$^{\rm f}$}&
\colhead{$L_{\rm X}$$^{\rm g}$}        &
\colhead{ID$^{\rm h}$}           }
\startdata
12 36 16.15 & +62 15 13.7&  130.4$^{+14.6}_{-13.3}$& 68.8$^{+10.2}_{-9.0}$& 57.4$^{+10.8}_{-9.3}$& $0.85^{+0.20}_{-0.18}$& $0.9^{+0.2}_{-0.2}$ & 1.01& 0.16& 0.80& 1.30$^{+0.02}_{-0.03}$ & $8.3^{+12.2}_{-5.2}$ & AGN\\

12 36 18.33 & +62 15 50.5$^{\rm i}$&  29.4$^{+6.9}_{-5.8}$ & $<$17.5 & $<$17.2 & $-$ & 1.4 & 0.17 & $<$0.04 & $<$0.19 & 1.41$^{+0.03}_{-0.03}$ & $0.4^{+0.6}_{-0.2}$ & U \\

12 36 20.3  & +62 17 01 & $<$15.5 & $<$10.4 & $<$14.1 &  &  & $<$0.10 & $<$0.03 & $<$0.19 & $>$1.49 & $-$ & U \\

12 36 21.1  & +62 12 50 & $<$11.1 & $<$5.3  & $<$10.3 &  &  & $<$0.07 & $<$0.01 & $<$0.14 & $>$1.50 & $-$ & U \\

12 36 22.65 & +62 16 29.7&  63.4$^{+11.3}_{-10.1}$& $<$13.2& 56.9$^{+10.7}_{-9.4}$& $>$4.39& $<$~--0.6& 1.03&   $<$0.03& 1.15 & 1.41$^{+0.03}_{-0.03}$ & $13.2^{+19.3}_{-8.3}$ & AGN \\

12 36 29.13 & +62 10 45.8&  198.4$^{+16.8}_{-15.7}$& 66.3$^{+10.0}_{-8.9}$& 134.7$^{+14.1}_{-13.0}$& $2.05^{+0.38}_{-0.33}$& $0.1^{+0.2}_{-0.2}$ & 2.32& 0.16& 2.26& 1.28$^{+0.03}_{-0.03}$ & $3.1$ & AGN \\

12 36 46.05 & +62 14 48.7&  13.8$^{+6.0}_{-4.8}$& 12.6$^{+6.8}_{-4.9}$& $<$7.5& $<$0.60& 1.4& 0.08& 0.03& $<$0.09& 1.49$^{+0.06}_{-0.04}$ & $0.4^{+0.6}_{-0.2}$ & U \\

12 36 52.06 & +62 12 25.7 & $<$9.8 & $<$10.8 & $<$4.0 &  &  & $<$0.06 & $<$0.03 & $<$0.05 & $>$1.48 & $<$1.1 & U \\

12 37 07.21 & +62 14 08.1 & 84.4$^{+11.4}_{-10.2}$& 32.6$^{+7.4}_{-6.2}$& 52.3$^{+9.4}_{-8.2}$& $1.62^{+0.48}_{-0.39}$& $0.3^{+0.3}_{-0.2}$ & 0.98& 0.09& 0.92& 1.34$^{+0.03}_{-0.03}$ & $15.9^{+24.5}_{-10.2}$ & AGN \\

12 37 12.07 & +62 12 11.3 & 30.7$^{+8.2}_{-7.0}$ & $10.5^{+5.3}_{-4.1}$ & $20.9^{+7.2}_{-6.0}$ & $2.00^{+1.45}_{-0.88}$ & $0.1^{+0.5}_{-0.5}$ & 0.39 & 0.03 & 0.38 & 1.44$^{+0.02}_{-0.02}$ & $>$11.3 & AGN \\ 

\enddata

\tablenotetext{a}{Source coordinates. Taken from Table~1.}
\tablenotetext{b}{Source counts and errors or upper limit. Determined following the procedure given in Brandt et~al. (2001) for sources detected with {\sc wavdetect} false-positive probability threshold of 10$^{-7}$ and from {\sc wavdetect} for sources detected with {\sc wavdetect} false-positive probability threshold of 10$^{-5}$. ``FB'' indicates full band, ``SB'' indicates soft band, and ``HB'' indicates hard band. Upper limits are calculated following \S3.2.1 of Brandt et~al. (2001).}
\tablenotetext{c}{Ratio of the count rates in the 2.0--8.0~keV and 0.5--2.0~keV bands. The errors were calculated following the ``numerical method" described in \S1.7.3 of Lyons (1991).}
\tablenotetext{d}{Effective photon index for the 0.5--8.0~keV band, calculated from the band ratio. The photon indices for the X-ray sources detected with a threshold of 10$^{-5}$ and those with a low number of counts have been set to $\Gamma =1.4$, a value representative of the X-ray background spectral slope; see Brandt et~al. (2001). The photon index is related to the energy index by $\alpha$~=~$\Gamma$-1 where $F_{\nu}\propto\nu^{-\alpha}$.}
\tablenotetext{e}{Fluxes are in units of $10^{-15}$~erg~cm$^{-2}$~s$^{-1}$. These fluxes have been calculated following the method described in Brandt et~al. (2001). They have not been corrected for Galactic absorption. ``FB'' indicates full band, ``SB'' indicates soft band, and ``HB'' indicates hard band.}
\tablenotetext{f}{Submm-to-X-ray spectral index ($\alpha_{\rm SX}$) calculated following equation 3.}
\tablenotetext{g}{Rest-frame full-band luminosity in units of $10^{43}$~erg~s$^{-1}$ calculated following equation 1. The uncertainties in the calculated luminosities are determined using the upper and lower redshift bounds. For the AGNs the unabsorbed luminosity is given (corrected assuming the absorption is Compton thin and the underlying unabsorbed emission is a $\Gamma=2.0$ power law), while for the other sources $\Gamma=2.0$ is assumed. We are unable to place luminosity constraints for the two X-ray-undetected sources with redshift lower limits.}
\tablenotetext{h}{Source classification based on X-ray properties; see \S3.2. ``AGN'' indicates an AGN and ``U'' indicates that good X-ray spectral constraints could not be placed.}
\tablenotetext{i}{X-ray counterpart detected with {\sc wavdetect} with false-positive probability threshold of 10$^{-5}$.}

\newpage

\end{deluxetable}

%
%

\begin{table}
\begin{center}
\normalsize
\caption{X-ray spectral fits for the AGN-classified sources}
\begin{tabular}{cccc}
\hline
\hline
Object name & PL & \multicolumn{2}{c}{PL+ABS} \\
\cline{3-4} \\
 & $\Gamma^{\rm a}$ & $\Gamma^{\rm b}$ & $N_{\rm H}^{\, \rm c}$~(cm$^{-2}$) \\
\hline

123616.1+621513 & $1.0^{+0.3}_{-0.3}$  & $2.2^{+0.8}_{-0.7}$ & $0.8^{+0.4}_{-0.4}$ \\
123622.6+621629 & $-0.7^{+0.4}_{-0.4}$ & $0.1^{+1.0}_{-0.8}$ & $1.4^{+1.8}_{-1.3}$ \\
123629.1+621045 & $0.3^{+0.2}_{-0.2}$  & $1.3^{+0.6}_{-0.5}$ & $1.0^{+0.6}_{-0.5}$ \\
123707.2+621408 & $0.5^{+0.3}_{-0.3}$  & $1.6^{+0.8}_{-0.7}$ & $1.1^{+0.9}_{-0.7}$ \\
123712.0+621211 & $0.0^{+0.5}_{-0.6}$  & $0.5^{+0.5}_{-1.2}$ & $<$0.8              \\

\hline
\end{tabular}
\vskip 2pt
\footnotesize
{\sc Note. ---} 
The uncertainties refer to the 90\% confidence level (for one interesting parameter).\\
$^{\rm a}$ Best-fit photon index for Galactic absorption model. \\
$^{\rm b}$ Best-fit photon index for absorbed power-law model.\\
$^{\rm c}$ Best-fit column density in units of $10^{22}$~cm$^{-2}$ at $z=0$; the column density at the source redshift ($N_{\rm H,z}$) is related to the best-fit column density at $z=0$ by $N_{\rm H,z}\approx(1+z)^{2.6}~N_{\rm H}$.\\
\end{center}
\end{table}
\normalsize

\end{document}